\documentstyle[12pt,epsfig]{article}
\begin{document}
\noindent
\begin{center}
{\Large {\bf On the Chameleon Brans-Dicke Cosmology}}\\ \vspace{2cm}
${\bf Yousef~Bisabr}$\footnote{e-mail:~y-bisabr@srttu.edu.}\\
\vspace{.5cm} {\small{Department of Physics, Shahid Rajaee Teacher
Training University,
Lavizan, Tehran 16788, Iran}}\\
\end{center}
\vspace{1cm}
\begin{abstract}
We consider a generalized Brans-Dicke model in which the scalar field has a potential function and is also allowed to couple
non-minimally with the matter sector.  We assume a power law form for the potential and the coupling functions as the inputs of the
model and show that acceleration of the universe can be realized for a constrained range of exponent of the potential function.  We also argue that this accelerating
phase is consistent with a large and positive Brans-Dicke parameter. In our analysis,
the potential plays a more important role with respect to the coupling function in dynamics of the universe as the latter does not contribute
to any of the relations characterizing evolution of scale factor of the universe and the scalar field. However, we will show that the coupling function
is closely related to magnitude and direction of the energy transfer between matter and the scale field.  We use this fact and some thermodynamic aspects of the model to put
some constraints on the
coupling function.  In particular, we argue that the second law of thermodynamics constrains direction of the overall energy transfer.

\end{abstract}
~~~~~~~PACS Numbers: 04.50.Kd, 04.20.Cv, 95.36.+x \vspace{3cm}
\section{Introduction}
Cosmological observations on expansion history of the universe can be interpreted as evidence either for existence of some exotic matter components or for
modification of the gravitational theory. In the first route of interpretation one can take a mysterious cosmic fluid with
sufficiently large and negative pressure, dubbed dark energy.  In the second route, however, one attributes the accelerating expansion to
a modification of general relativity. Such modified gravity models can be obtained in different ways. For instance, one can replace the Ricci scalar in the Einstein-Hilbert action by some functions $f(R)$ (for a review see, e.g., \cite{1} and references therein), or by considering a scalar partner for the metric tensor for
describing geometry of spacetime, the so-called scalar-tensor gravity.  The prototype of the latter approach is Brans-Dicke (BD)
theory \cite{BD} which its original motivation was the search for a theory containing Mach's principle.  As the simplest and best-studied generalization of general relativity, it is natural to think about the BD scalar field as a possible candidate for producing cosmic acceleration without invoking auxiliary fields or exotic matter systems. In fact, there have been many attempts to show that BD model can potentially explain the cosmic acceleration. It is shown that this theory can actually produce a non-decelerating expansion for low negative values of the BD parameter $\omega$ \cite{ban}. Unfortunately, this conflicts with the lower bound imposed on this parameter by solar system experiments \cite{will}.  Due to this difficulty, some authors propose modifications of the BD model such as introducing some potential functions for the scalar field \cite{ban1}, or considering a field-dependent BD parameter \cite{ban2} without resolving the problem.\\
In a general scalar-tensor theory there is a non-minimal coupling between the scalar field and Ricci scalar while the former minimally
couples with the matter sector.  In other terms, there is no an explicit coupling between the scalar field and matter systems in Jordan frame representation.  In BD theory, in its original form, the motivation for such a minimal coupling was to keep the theory in accord with weak equivalence
principle \cite{BD}.   However, there has been recently a tendency in the literature to go a step further and consider a non-minimal coupling between the scalar field and matter systems as well by introducing an arbitrary function of the scalar field as a coupling function.  In these models, the scalar field is regarded as a chameleon field as it can be heavy enough in the environment of the laboratory tests so that the local gravity constraints are satisfied. Meanwhile, it can be light enough in the low-density cosmological environment to be considered as a candidate for dark energy.  Such a chameleon-matter coupling was first introduced in BD model to achieve accelerating expansion of the universe for sufficiently large BD
parameter \cite{ban3}.  It is shown that even though absolute value of $\omega$ is enlarged due to such a non-minimal coupling, its negative sign makes this theory remain inconsistent
with observations.  Later works apply non-minimal coupling to general scalar-tensor theories. For instance, \cite{fan} considers stability analysis and possibility
of phantom crossing with an assumption that potential of the scalar field and the coupling function have power-law forms, or bouncing solutions and some cosmological tests are investigated
in \cite{fan1}.    \\
In the present work, our primary interest is to consider the possibility that such an anomalous coupling of matter systems can produce accelerating expansion of the universe
for a sufficiently large positive BD parameter.  As a first step, we will study conservation laws in such models in section $2$.  By writing the energy balance and the
geodesic equations, we will show that how they are modified in terms of the introduced coupling function.  It is also shown that
the non-conservation of matter energy density means energy transfer between matter and the scalar field with a constant rate.  In section $3$, we
study this model in a cosmological setting.
We assume a power law form for the potential and the coupling function and show that accelerating expansion of the universe can be realized within a class of power law
solutions of the field equations.  In this class of solutions $\omega$ can take positive and large values.  We also discuss some thermodynamic aspects of the model and show that
the overall energy transfer should be into the matter system if the second law of thermodynamics is to be fulfilled.  We use this fact and some recent observations to tightly constrain
the exponent of the coupling function.
~~~~~~~~~~~~~~~~~~~~~~~~~~~~~~~~~~~~~~~~~~~~~~~~~~~~~~~~~~~~~~~~~~~~~~~~~~~~~~~~~~~~~~~~~~~~~~~~~~~~~
\section{The model}
We consider the action functional
\begin{equation}
S=\frac{1}{2}\int d^4x \sqrt{-g} \{\phi R-\frac{\omega}{\phi}g^{\mu\nu}\nabla_{\mu}\phi \nabla_{\nu}\phi -2V(\phi)+2f(\phi)L_m\}
\label{1}\end{equation}
where $R$ is the Ricci scalar, $\phi$ is the BD scalar field, $V(\phi)$ and $f(\phi)$ are some analytic functions.  Here
the matter Lagrangian density, denoted by $L_m$, is coupled with $\phi$ via the function $f(\phi)$. This allows a non-minimal coupling between the matter system and $\phi$, thus the latter appears as a chameleon field.  Taking $f(\phi)=1$, we return to the BD action with
a potential function $V(\phi)$.  \\
Varying the action with respect to the metric $g_{\mu\nu}$ and $\phi$ yields
the field equations, given by,
\begin{equation}
\phi G_{\mu\nu}=T^{\phi}_{\mu\nu}+f(\phi)T^{m}_{\mu\nu}
\label{2}\end{equation}
\begin{equation}
(2\omega+3)\Box \phi+2(2V-V'\phi)=T^m f-2f'\phi L_m
\label{3}\end{equation}
where prime indicates differentiation with respect to $\phi$, $T^m=g^{\mu\nu}T^m_{\mu\nu}$ and
\begin{equation}
T^{\phi}_{\mu\nu}=\frac{\omega}{\phi}(\nabla_{\mu}\phi \nabla_{\nu}\phi-\frac{1}{2} g_{\mu\nu}\nabla_{\alpha}\phi \nabla^{\alpha}\phi)
+(\nabla_{\mu}\nabla_{\nu}\phi-g_{\mu\nu}\Box \phi)-V(\phi)g_{\mu\nu}
\label{4}\end{equation}
\begin{equation}
T^m_{\mu\nu}=\frac{-2}{\sqrt{-g}}\frac{\delta (\sqrt{-g}L_m)}{\delta g^{\mu\nu}}
\label{5}\end{equation}
Due to the explicit coupling of the matter system with $\phi$, the stress tensor $T^m_{\mu\nu}$ is not divergence free.
This can be seen by applying the Bianchi identities $\nabla^{\mu}G_{\mu\nu}=0$ to (\ref{2}), which leads to
\begin{equation}
\nabla^{\mu}T^m_{\mu\nu}=(L_mg_{\mu\nu}-T^m_{\mu\nu})\nabla^{\mu}\ln f
\label{6}\end{equation}
As it is clear from (\ref{6}), details of the energy exchange between matter and $\phi$ depends on the explicit form
of the matter Lagrangian density $L_m$.  Here we consider a perfect fluid energy-momentum tensor as a matter system
with energy density $\rho_m$ and pressure $p_m$. \\
There are different choices
for the perfect fluid Lagrangian density which all of them leads to the same energy-momentum tensor and field equations in the context
of general relativity \cite{2} \cite{3}.  The two Lagrangian densities that have been widely used in the literature are
$L_m=p_m$ and $L_m=-\rho_m$ \cite {3a} \cite{4} \cite{5}.  For a perfect fluid that does not couple
explicitly to $\phi$ (i.e., for $f(\phi) = 1$), the two Lagrangian densities $L_m =p_m$ and $L_m=-\rho_m$ are perfectly
equivalent, as discussed in \cite{4} \cite{5}. However, in the model presented here the expression of $L_m$ enters explicitly the field equations
and all results strongly depend on the choice of $L_m$.  In fact, it is shown that there is a strong debate about equivalency of different expressions attributed to the Lagrangian density of a coupled perfect fluid \cite{6}.  \\
Here we consider $L_m=p_m$.  In this case, although fluid elements follow geodesics of the background metric and there
is no additional force, the matter is still non-conserved \cite{bisabr}
\begin{equation}
\dot{\rho}_m+3H(\gamma+1)\rho_m=-(\gamma+1)\rho_m\frac{\dot{f}}{f}
\label{b8}\end{equation}
This has the solution
\begin{equation}
\rho_m=\rho_0 a^{-3(\gamma+1)}f^{-(\gamma+1)}
\label{b13}\end{equation}
with $\rho_0$ being an integration constant.  This indicates that evolution of matter density strongly depends on the coupling
function. When $f=1$, (\ref{b13}) reduces to the standard evolution law for the matter energy density.  However, as we have already stated, the coupling function in this framework does not affect a matter system with $\gamma=-1$ (a cosmological constant) as it is clear from (\ref{b13}).

~~~~~~~~~~~~~~~~~~~~~~~~~~~~~~~~~~~~~~~~~~~~~~~~~~~~~~~~~~~~~~~~~~~~~~~~~~~~~~~~~~~~~~~~~~~~~~~~~~~~~~~~~~~~~~~~~~~~~~~~~~~~~~~~~~~~

\section{A cosmological setting}
 Now we apply the above framework to a homogeneous and isotropic universe described by the metric (\ref{b3}).
 In a spatially flat case, the equations (\ref{2}) and (\ref{3}) become
\begin{equation}
3\frac{\dot{a}^2}{a^2}=\frac{f}{\phi}\rho_m+\frac{1}{2}\omega \frac{\dot{\phi}^2}{\phi^2}-3H\frac{\dot{\phi}}{\phi}+\frac{V}{\phi}
\label{c1}\end{equation}
\begin{equation}
3\frac{\ddot{a}}{a}=-\frac{3\rho_m}{\phi(2\omega+3)}[\gamma \phi f'+f(\omega(\gamma+\frac{1}{3})+1)]-\omega\frac{\dot{\phi^2}}{\phi^2}
+3H\frac{\dot{\phi}}{\phi}+\frac{1}{(2\omega+3)}[3V'+(2\omega-3)\frac{V}{\phi}]
\label{c3}\end{equation}
\begin{equation}
(2\omega+3)(\ddot{\phi}+3H\dot{\phi})-2(2V-\phi V')=(1-3\gamma)f\rho_m+2\gamma\phi f'\rho_m
\label{c4}\end{equation}
 To proceed further, we choose
 $V(\phi)=V_0\phi^{l_1}$ and $f(\phi)=f_0\phi^{l_2}$ in which $l_1$ and $l_2$ are constant parameters and $V_0$ and $f_0$ are positive quantities.  We assume power law forms for evolution of the scale factor and the scalar field
\begin{equation}
a(t)=a_0t^n
\label{cc1}\end{equation}
\begin{equation}
\phi(t)=\phi_0t^m
\label{cc2}\end{equation}
Inserting these ansatz functions together with (\ref{b13}) into (\ref{c1})-(\ref{c4}) gives the parameters
$n$ and $m$ in terms of $l_1$ and $l_2$.  The latter parameters should be regarded as the inputs
of the model described by (\ref{1}) as they characterize the shape of the functions $V(\phi)$ and $f(\phi)$.  \\Since the left hand side of (\ref{c1})-(\ref{c4})
falls as $t^{-2}$, we arrive at the following relationships
\begin{equation}
m(l_1-1)=-2
\label{c5}\end{equation}
\begin{equation}
m(l_2-1)-(3n+ml_2)(\gamma+1)=-2
\label{c6}\end{equation}
The equations (\ref{c1}) and (\ref{c4}), give then the consistency relations
\begin{equation}
3n^2+3mn-\frac{1}{2}m^2\omega-V_0\phi_0^{l_1-1}-f_0^{-\gamma}\rho_0a_0^{-3(\gamma+1)}\phi_0^{-(\gamma l_2+1)}=0
\label{c7}\end{equation}
\begin{equation}
(2\omega+3)(m(m-1)+3mn)-2(2-l_1)V_0\phi_0^{l_1-1}+[(3-2l_2)\gamma-1]f_0^{-\gamma}\rho_0a_0^{-3(\gamma+1)}\phi_0^{-(\gamma l_2+1)}=0
\label{c8}\end{equation}
Combining (\ref{c5}) and (\ref{c6}) gives
\begin{equation}
n=\frac{2(\gamma l_2+l_1)}{3(\gamma+1)(l_1-1)}
\label{c9-1}\end{equation}
For $\gamma\neq 0$, the conditions for the universe to be expanding ($n>0$) and accelerating ($n>1$) are the following\footnote{We assume that the matter system satisfies the
weak energy condition $\gamma+1>0$.}
$$
n>0~~\rightarrow~~ l_2>-l_1/\gamma~~~~~~~~~~~~~~l_1>1~~~~~~~~~~~~~~~~~~~~~~~~~~~~~~~~~~~~
$$
\begin{equation}
~~~~~~~~~~~~~~~~l_2<-l_1/\gamma~~~~~~~~~~~~~~l_1<1~~~~~~~~~~~~~~~~~~~~~~~~~~~~~~~~~~~
\label{c9-2}\end{equation}
$$
n>1~~\rightarrow ~~l_2>\frac{3}{2}(l_1-1)+\frac{1}{2\gamma}(l_1-3)~~~~~~~~~l_1>1~~~~~~~~~~~~~~~~~~~
$$
\begin{equation}
~~~~~~~~~~~~~~~~~l_2<\frac{3}{2}(l_1-1)+\frac{1}{2\gamma}(l_1-3)~~~~~~~~~l_1<1~~~~~~~~~~~~~~~~~~~
\label{c9-3}\end{equation}
As an illustration, let us consider the potential $V(\phi)=V_0 \phi^2$ in a radiation-dominated universe.  In this
case, $\gamma=1/3$ and expansion of the universe implies $l_2>-6$. On the other hand, the success of big bang nucleosynthesis gives us a strong evidence of the radiation-dominated
decelerated phase which leads to $l_2<0$.  Thus a scalar field with a quadratic potential which couples with radiation requires that $-6<l_2<0$.  Note that the relation (\ref{c5}) indicates that for $l_1>1$ the scalar field is a decreasing function of time ($m<0$).  This implies that the coupling function
$f\propto t^{ml_2}$
increases with time or the matter-chameleon coupling gets stronger as the universe expands.  \\
This feature is changed in a matter-dominated universe.  For $\gamma=0$, the expression (\ref{c9-1}) reduces to
\begin{equation}
n=\frac{2l_1}{3(l_1-1)}
\label{c9}\end{equation}
This relation indicates that $n>1$ can not be realized if $l_1<0$.  On the other hand, when $l_1>0$ accelerating expansion of the universe requires that
\begin{equation}
1<l_1<3
\label{c99}\end{equation}
The parameter $l_2$ does not enter the above condition and it seems that the scalar field effectively decouples from the matter system in a matter-dominated
universe. Exploring the
relations (\ref{c5}) and (\ref{c6}) reveals that $l_2$ also disappears from these relation when we set $\gamma=0$.  The same is true
for the consistency relations (\ref{c7}) and (\ref{c8}).  \\
As it is evident, the above conditions (\ref{c9-2}), (\ref{c9-3}) and (\ref{c99})
do not put any constraint on the BD parameter $\omega$.  To obtain such a constraint, one should consider  the right hand side of (\ref{c3}) as a function
\begin{equation}
C=-\frac{3\rho_m}{\phi(2\omega+3)}[\gamma \phi f'+f(\omega(\gamma+\frac{1}{3})+1)]-\omega\frac{\dot{\phi^2}}{\phi^2}
+3H\frac{\dot{\phi}}{\phi}+\frac{1}{(2\omega+3)}[3V'+(2\omega-3)\frac{V}{\phi}]
\label{c10}\end{equation}
and establish the condition for having $C>0$.  For $\gamma=0$, the condition is
\begin{equation}
V_0\phi_0^{l_1-1}[\frac{2\omega+3(l_1-1)}{2\omega+3}]-\frac{\rho_0 a_0^{-3}(\omega+3)}{\phi_0(2\omega+3)}-m^2\omega-\frac{6n}{l_1-1}>0
\label{c11}\end{equation}
At this point, attention is focused on the question that whether it is possible to have accelerated expansion for a large
positive $\omega$.  To answer the question, let us consider the above expression for $\omega>>1$,
\begin{equation}
V_0\phi_0^{l_1-1}-\frac{\rho_0 a_0^{-3}}{2\phi_0}-\frac{6n}{l_1-1}>m^2\omega
\label{c11-1}\end{equation}
If the left hand side effectively takes a negative sign then the inequality can not be satisfied for a positive BD parameter.  To make an
estimation, we first note that $\phi_0^{-1}\sim G\sim M_p^{-2}$ and $\rho_0 a_0^{-3}\sim M$ with $G$, $M_p$ and $M$ being respectively the gravitational constant, the Planck mass and the total mass content in the
universe. Then we can rewrite (\ref{c11-1}) in the form
\begin{equation}
V_0 M_p^{2(l_1-1)}-\frac{M}{2M_p^2}-\frac{6n}{l_1-1}>m^2\omega
\label{c11-2}\end{equation}
It is an observational fact that matter density of the universe is of the same order of the critical density $\rho_c=3H^2M_p^2/8\pi $ with $H$ being the Hubble parameter.  This leads to a relation between $M$ and $H$ such that $M/M_p^2\sim R$, which $R\sim H^{-1}$ is the Hubble radius\footnote{One can say that the radius of the universe coincides with its Schwarzschild radius $2GM$ \cite{sa}.}.
Substituting this result into (\ref{c11-2}), gives
\begin{equation}
M_p^{2(l_1-1)}(V_0-\frac{R^{l_1}}{M^{l_1-1}})-\frac{6n}{l_1-1}>m^2\omega
\label{c11-3}\end{equation}
Using $R\sim 10^{26}(meter)$ and $M\sim 10^{96}(meter)^{-1}$\footnote{We use the unit system in which $\hbar=c=1.$}, one can see that
 the term containing $V_0$ is a leading term on the left hand side of (\ref{c11-3}) unless $l_1$ takes values near
the lower bound in (\ref{c99}), namely $l_1\approx 1$.  In other words, when $l_1$ is not close to unity one can write $\omega<M_p^{2(l_1-1)}V_0/m^2$ which indicates that the BD parameter can
take positive large values.  Note that there is no need for fine-tuning of the constant $V_0$ since the upper bound given by the latter relation is
sufficiently large due to the appearance of $M_p$.  It can be easily checked that for $l_1\approx 1$, $R^{l_1}/M^{l_1-1}\approx R\sim 10^{26}(meter)$
and the left hand side of (\ref{c11-3}) take a negative sign.  In this case, the inequality can not be satisfied for $\omega>0$.
It is interesting to note that the coupling function, or the parameter $l_2$, does not play any role in the above argument and it is only
the potential function that is relevant. As we have already seen, this is also the case (for $\gamma=0$) when one enters the arguments
concerning the expansion of the universe in matter-dominated era.\\
Despite the irrelevant role of the coupling function in the latter arguments, it affects the dynamics of the matter energy density.  Combining (\ref{cc1}) and (\ref{cc2}) with (\ref{b13}) gives
\begin{equation}
\rho_m \propto a^{-3(\gamma+1)+\varepsilon}
\label{c12}\end{equation}
where
\begin{equation}
\varepsilon=-\frac{m}{n}\l_2(\gamma+1)
\label{c13}\end{equation}
In fact, the parameter $\varepsilon$ measures the modification of matter expansion rate due to its interaction with the scalar field.  The relation (\ref{c12}) states that when $\varepsilon>0$ matter is created and energy is constantly injecting into the matter so that the latter will dilute more slowly compared to its standard evolution $\rho_m \propto a^{-3(\gamma+1)}$.  Similarly, when $\varepsilon<0$ the reverse is true, namely that matter is
annihilated and direction of the energy transfer is outside of the matter system so that the rate of the dilution is faster than the standard one.\\
We may use (\ref{c6}) to write (\ref{c13}) in the form
\begin{equation}
n=\frac{2l_2 (\gamma+1)}{3l_2(\gamma+1)^2-\varepsilon(\gamma\l_2+1)}
\label{c14}\end{equation}
When $m=0$, $\phi$ takes a constant configuration which is given by $\phi_0\sim G^{-1}$ and the model
(\ref{1}) reduces to Einstein gravity with a cosmological term.  This case corresponds to $\varepsilon=0$ and the solution (\ref{c14})
reduces to $n=2/3(\gamma+1)$, the case in the standard cosmology.\\
For $\gamma=0$, the expression (\ref{c14}) takes the form
\begin{equation}
n=\frac{2}{3-\frac{\varepsilon}{l_2}}
\label{c14-1}\end{equation}
One can use this relation to constrain the parameters $\varepsilon$ and $l_2$.  We first note that from the modified matter expansion rate (\ref{c12}) we should expect $|\varepsilon|<<1$ since so far there has been
no report from observations about an anomalous expansion rate. In other terms, if there exits any deviation from the standard evolution of matter
density it must be very small. Taking $\varepsilon$ as an independent parameter, one can compare the modified expansion rate
with observations and perform data fitting to estimate $|\varepsilon|$.  In fact, it is shown that cosmological observations are consistent with $|\varepsilon|<0.1$ \cite{wang} \cite{al} \cite{a2}.  This result puts an upper bound on absolute value of the parameter $l_2$.  The condition for
accelerating expansion inferred by (\ref{c14-1}) is  $\varepsilon/l_2>1$  which leads to $|l_2|<0.1$.\\
On the other hand, the expression (\ref{c14-1}) implies that the parameters $\varepsilon$ and $l_2$ should have the same sign in order that $n>1$.  Thus direction of the energy transfer between the matter system and the scalar field is characterized either by $\varepsilon$ or
$l_2$.  Following \cite{al}, we argue that $\varepsilon>0$ as required by the second law of thermodynamics.  To do this, we should investigate some thermodynamic features
of the matter-chameleon coupling described by (\ref{1}).  A thermodynamic description of a perfect fluid matter system requires the knowledge of the particle flux $N^{\alpha}=nu^{\alpha}$
and the entropy flux $S^{\alpha}=n\sigma u^{\alpha}$ where $n = N/a^3$ and $\sigma = S/N$ are, respectively, the concentration and the specific entropy (per particle) of the created
or annihilated particles.  Since the energy density of matter is given by $\rho_m=nM$ with $M$ being the mass of each particle, the appearance of the extra term in
the energy balance equation (\ref{b8}) means that this extra-change of $\rho_m$ can be attributed to a change of $n$ or $M$.  Here we assume that the mass of each matter particle remains
constant and the extra term in the energy balance equation only leads to a change of the number density $n$. In this case, the equations (\ref{b8}) can
be written as ($\gamma=0$)
\begin{equation}
\dot{n}+3Hn=n\Gamma
\label{c15}\end{equation}
where
\begin{equation}
\Gamma\equiv-\frac{\dot{f}}{f}=\varepsilon\frac{\dot{a}}{a}
\label{c16}\end{equation}
 is the decay rate.  We also assume that the overall energy transfer is an adiabatic processes in which matter particles are continuously created or annihilated while
 the specific entropy per particle remains constant during the whole processes ($\dot{\sigma}=0$) \cite{li}.  This means that
 \begin{equation}
 \frac{\dot{S}}{S}=\frac{\dot{N}}{N}=\Gamma
 \label{c17}\end{equation}
From $n\propto a^{-3+\varepsilon}$, we can see that the total number of particles scales as $N\propto a^{\varepsilon}$, and the second law of thermodynamics
$\dot{S}\geq 0$ implies that $\varepsilon\geq0$.  This conclusion can also be drawn by (\ref{c16}) since $\Gamma\geq0$ requires that $\varepsilon\geq0$ in an
expanding universe.  Hence, the chameleon scalar field should suffer energy reduction and the matter system should gain energy during expansion of the universe if the second law of thermodynamics is to be fulfilled.
~~~~~~~~~~~~~~~~~~~~~~~~~~~~~~~~~~~~~~~~~~~~~~~~~~~~~~~~~~~~~~~~~~~~~~~~~~~~~~~~~~~~~~~~~~~~~~~~~~~~~~~~~~~~~~~~~~~~~~~~~~~~~~~~~~~~~~~
\section{Conclusions}
In this work we have studied some features of the generalized BD model in which the scalar field is allowed to couple non-minimally
with matter sector.  The matter expansion law and the geodesic equation have modified due to this non-minimal coupling.  The modification depends on the choice of $L_m$, Lagrangian density of the matter system. For instance, in our choice matter conservation equation is modified
while there is no extra force in the geodesic equation since $L_m = p_m$ leads to
vanishing of the first term on the right hand side of the equation (\ref{b7}). \\We have found power law solutions for the scale factor and the scalar field. In this class of solutions,  when $\gamma=0$ accelerating expansion of the universe can be realized for $1<\l_1<3$.  Our analysis also
indicates that this accelerating phase is consistent with a positive and large BD parameter.\\ The fact that the parameter $l_2$ does not contribute to the condition for cosmic speed-up
 in the case of $\gamma= 0$, may be regarded as a signal for an irrelevant role of the function $f(\phi)$ in the matter-dominated era.
However, it is evident from (\ref{b13}) that $f(\phi)$ is important in the evolution of matter density.  We have reformulated the matter expansion law as $\rho_m \propto a^{-3(\gamma+1)+\varepsilon}$ where the parameter $\varepsilon$ characterizes both magnitude and direction of
the energy transfer.  We have argued that such a reformulation has two important consequences.  First, since the magnitude of $\varepsilon$ can be fixed by observation (as it is recently suggested that $|\varepsilon|<0.1$) accelerating expansion which requires that $l_2<\varepsilon$ sets an upper bound on the absolute value of $l_2$.  Second, the sign of $\varepsilon$
is restricted by the second law of thermodynamics to assume only positive values so that the direction of energy transfer is into the matter system.  This constrains the exponent
of the coupling function to take values within the range $0<l_2<0.1$.


\begin{thebibliography}{99}
\bibitem{1}T. P. Sotiriou and V. Faraoni, Rev. Mod. Phys. {\bf 82}, 451 (2010)\\
S. Nojiri and S. D. Odintsov, Phys. Rept. {\bf 505}, 59 (2011)
\bibitem{BD}C. Brans and R. H. Dicke, Phys. Rev. {\bf 124}, 925 (1961)
\bibitem{ban}N. Banerjee and D. Pavon, Phys. Rev. D {\bf 63}, 043504 (2001)
\bibitem{will}C.M. Will, Theory and Experiment in Gravitational Physics, (Cambridge University Press, 3rd edition, Cambridge, 1993)\\
C. M. Will, Liv. Rev. Rel. {\bf 9}, 3 (2005)
\bibitem{ban1}O. Bertolami and P. J. Martins, Phys. Rev. D {\bf 61}, 064007 (2000)\\
M. K. Mak and T. Harko, Europhys. Lett. {\bf 60}, 155 (2002)
\bibitem{ban2}W. Chakraborty and U. Debnath, arxiv:0807.1776v1
\bibitem{ban3}S. Das and N. Banerjee, Phys. Rev. D {\bf 78}, 043512 (2008)
\bibitem{fan}H. Farajollahi and A. Salehi, JCAP {\bf 1011}, 006 (2010)\\
H.Farajollahi and A. Salehi, JCAP {\bf 07}, 036 (2011)
\bibitem{fan1}H. Farajollahi, M. Farhoudi, A. Salehi and H. Shojaie, Astrophys. Space Sci. {\bf 337}, 415 (2012)
\bibitem{2}B. Schutz,  Phys. Rev. D {\bf 2} 2762 (1970)\\
J. D. Brown,  Class. Quant. Grav. {\bf 10}, 1579 (1993)
\bibitem{3}S. W. Hawking and G. F. R. Ellis, \emph{The Large Scale Structure of Space-Time} (Cambridge 1973, Cambridge University Press)
\bibitem{3a}O. Bertolami, C. G. Bohmer, T. Harko and F. S. N. Lobo, Phys. Rev. D {\bf 75}, 104016 (2007)
S. Nojiri, S. D. Odintsov and P. V. Tretyakov, Prog. Theor. Phys. Suppl. {\bf 172}, 81 (2008)
\bibitem{4}
O. Bertolami and J. Paramos, arXiv:1003.1875v1\\
O. Bertolami and A. Martins, arXiv:1110.2379\\
O. Bertolami, P. Frazao and J. Paramos, Phys. Rev. D {\bf 83}, 044010 (2011)
\bibitem{5}T. P. Sotiriou and V. Faraoni, Class. Quant. Grav. {\bf 25}, 205002 (2008)
\bibitem{6}V. Faraoni, Phys. Rev. D {\bf 80}, 124040 (2009)
\bibitem{bisabr}Y. Bisabr, Phys. Rev. D {\bf 86}, 044025 (2012)
\bibitem{wang}P. Wang and  X. Meng, Class. Quant. Grav. {\bf 22}, 283 (2005)
\bibitem{al}
J. S. Alcaniz and  J. A. S. Lima, Phys. Rev. D {\bf 72}, 063516 (2005)
\bibitem{a2}
F. E. M. Costa, J. S. Alcaniz and J. M. F. Maia, Phys. Rev. D {\bf 77}, 083516 (2008)\\
J. F. Jesus, R. C. Santos, J. S. Alcaniz and J. A. S. Lima, Phys. Rev. D {\bf 78} 063514 (2008)\\
F. E. M. Costa and J. S. Alcaniz, Phys. Rev. D {\bf 81}, 043506 (2010)
\bibitem{sa}H. Salehi and Y. Bisabr, Int. J. Theo. Phys. {\bf 39}, 1241 (2000)\\
Y. Bisabr and H. Salehi, Class. Quant. Grav. {\bf 19}, 2369 (2002)
\bibitem{li}J. A. S. Lima, Phys. Rev. D {\bf 54}, 2571 (1996)








\end{thebibliography}
\end{document}